\documentclass{emulateapj}

\usepackage[export]{adjustbox}

\usepackage{color}
\newcommand{\CH}[1]{{\bf\color{blue} #1}}
\usepackage{soul}   %% use \st{}
\begin{document}
\title{HOROLOGIUM II: A SECOND ULTRA-FAINT MILKY WAY SATELLITE IN THE HOROLOGIUM CONSTELLATION}
\author{Dongwon Kim} 
\author{Helmut Jerjen} 

\affil{Research School of Astronomy and Astrophysics, The Australian National University, Mt Stromlo Observatory, via Cotter Rd, 
Weston, ACT 2611, Australia}

\email{dongwon.kim@anu.edu.au}

\begin{abstract}
We report the discovery of a new ultra-faint Milky Way satellite candidate, Horologium\,II, detected in the Dark Energy Survey Y1A1 public data. Horologium\,II features a half light radius of $r_{h}=47\pm10$\,pc and a total luminosity of $M_{V}=-2.6^{+0.2}_{-0.3}$ that place it in the realm of ultra-faint dwarf galaxies on the size-luminosity plane. The stellar population of the new satellite is consistent with an old ($\sim13.5$\,Gyr) and metal-poor ([Fe/H]$\sim-2.1$) isochrone at a distance modulus of $(m-M)=19.46\pm0.20$, or a heliocentric distance of $78\pm8$\,kpc, in the color-magnitude diagram. Horologium\, II has a distance similar to the Sculptor dwarf spheroidal galaxy ($\sim82$\,kpc) and the recently reported ultra-faint satellites Eridanus\,III ($87\pm8$\,kpc)  and Horologium\,I ($79\pm8$\,kpc). All four satellites are well aligned on the sky, which suggests a possible common origin. As Sculptor is moving on a retrograde orbit within the Vast Polar Structure when compared to the other classical MW satellite galaxies including the Magellanic Clouds, this hypothesis can be tested once  proper motion measurements become available.

\end{abstract}

\section{Introduction}

Over the last decades wide-field  imaging surveys have systematically revealed new satellite companions to the Milky Way (MW). 
Especially, the Sloan Digital Sky Survey~\citep[SDSS;][]{York2000} was instrumental in establishing a new class of stellar systems, the 
ultra-faint dwarf (UFD) galaxies ~\citep[e.g.][]{Willman2005,Zucker2006,Belokurov2006,Irwin2007,Walsh2007,Grillmair2009}, thereby more 
than doubling the number of known MW satellite galaxies over half the northern hemisphere. Deep imaging follow-ups and spectroscopic studies 
suggest that the UFDs hold typically old~\citep[e.g.][]{Munoz2010,Sand2012,Brown2014} and metal poor~\citep[e.g.][]{Kirby2008,Frebel2010,Norris2010} stellar 
populations. The high mass-to-light ratios of the UFDs ($M/L_{V}>100$) inferred from internal kinematics~\citep[e.g.][]{Martin2007,Simon2007,Simon2011} is one of properties that differentiate them from star clusters~\citep{Willman2012}. 
The efforts to find new MW satellites over a larger area of sky continued with the VST ATLAS~\citep{Shanks2015} and Pan-STARRS $3\pi$ (K. Chambers et al., in preparation) surveys, 
both of which have delivered a couple of discoveries to date~\citep{Belokurov2014,Laevens2014,Laevens2015}. Most recently, systematic searches based on the first data release (Y1A1) of the Dark Energy Survey~\cite[DES;][]{DES}, have continued the success of its predecessor the SDSS, 
unveiling nine new objects over $\sim1800$ square degrees in the southern sky~\citep{Koposov2015a, Bechtol2015}\footnote{We note that the satellite 
reported as Indus\,I by \cite{Koposov2015a} and as DES J2108.8-5109 by \cite{Bechtol2015} is identical to Kim\,2 that was discovered earlier by~\cite{Kim2015a}.}, some of which have been already confirmed as UFDs by spectroscopic investigations~\citep{Simon2015,Walker2015,Koposov2015b}. Other independent surveys, such as the Stromlo Missing Satellite Survey~\citep{SMS} and the Survey of the Magellanic Stellar History (SMASH; D. Nidever et al., in preparation), also took advantage of the power 
of the Dark Energy Camera (DECam) to boost the census of MW companions in the southern sky~\citep{Kim2015a,Martin2015}.   

The use of different detection algorithms also contributed significantly to the increase in the number of known MW satellites~\citep{Koposov2008,Invisibles}. 
Due to their extremely low surface brightness~\citep{Martin2008}, UFDs would be difficult to characterised without the help of such specialised data mining techniques. 
Further improvements to the detection sensitivity even led to new discoveries of stellar systems hiding in the pre-existing SDSS data~\citep[e.g.][]{Kim1,Kim2015b}.

Here we announce the discovery of a new ultra-faint MW satellite galaxy candidate found in the DES Y1A1 data. We note that this object Horologium\,II (Hor\,II) does not correspond to any object in the previous studies by \cite{Koposov2015a}  and \cite{Bechtol2015} or in catalogs including the NASA/IPAC Extragalactic Database and SIMBAD.

\section{Data Reduction and Discovery}

DES is a deep photometric survey using the wide-field ($\sim3$ square degree) Dark Energy Camera (DECam) imager that consists of 62 2k $\times$ 4k CCD chips installed at the 4-m Blanco Telescope located at Cerro Tololo Inter-American Observatory (CTIO). DES started operation in August 2013 and will cover $\sim5000$ square degrees of the Southern Sky in the vicinity of the Magellanic Clouds in five photometric bands ($grizY$) over five years.  The data used in this paper is its first year public data set, DESDM Y1A1, collected between August 2013 and February 2014 over approximately 1800 square degrees and released to the public by the NOAO Science archive after a one year proprietary period. This data set includes individual images and corresponding weight-maps processed by the DES data management (DESDM) pipeline. Each image is a 90s single exposure.  The instCal images we used for our analysis are bias, dark and flat-field corrected and contain the world coordinate system provided by the DESDM image processing pipeline~\citep[see][for more details]{Desai2012,DESDM}. 

We downloaded all the Y1A1 instCal images and corresponding weight-maps for the $g$ and $r$ bands from the NOAO Science archive using its SQL interface. Crossmatching the central coordinates of the images within $1\farcm0$ radius yielded 1980 image pairs between the two photometric bands. To produce photometric catalogs, we performed PSF photometry over the images using SExtractor/PSFEx~\citep{SExtractor,PSFEx} on a local 16 nodes/128 core computer cluster. We carried out star/galaxy separation based on the threshold $\mathtt{\left| SPREAD\_MODEL\right|<0.003+SPREADERR\_MODEL}$ as described in \cite{Koposov2015a}. The catalogs, which contained the instrumental magnitudes of the star-like objects, were crossmatched between $g$ and $r$ bands by employing STILTS~\citep{STILTS} with a $1\arcsec$ tolerance. We then calibrated the instrumental magnitudes of the star-like point sources with respect to the APASS DR\,8 stars by means of 500 bootstrap samples with 3-sigma clipping. On average, we found $\sim370$ crossmatches between the instrumental and APASS catalogs on each frame, which yielded photometric zero points with uncertainties $\sim0.003$ magnitudes in both $g$ and $r$ bands. The calibrated magnitudes were finally corrected for Galactic extinction using the reddening map by~\cite{Schlegel1998} and the correction coefficients from~\cite{Schlafly2011}.

We then applied our overdensity detection algorithm to the star catalogs to search for MW satellites. Briefly, this algorithm, following the approach by~\cite{Invisibles}, involves a photometric filtering process in which an isochrone mask is applied to select a single age/metallicity stellar population at a fixed distance modulus.  The density map generated from the selected stars is then convolved with a Gaussian kernel. The significance of local stellar overdensities is measured by comparing their signal-to-noise ratios to the smoothed density map. These steps are repeated shifting the isochrone mask over a range of distance moduli. The detection process is described in more details in ~\cite{Kim1}. As part of this photometric analysis of the entire 1800 sqr deg of $gr$ images of the Y1A1 data set, we successfully recovered all the UFD candidates reported by \cite{Koposov2015a} and \cite{Bechtol2015}; e.g. Phoenix\,II ($17\sigma$), Pictoris\,I ($13\sigma$), Tucana\,II ($11\sigma$), Eridanus\,III ($10\sigma$), and Grus\,I ($9\sigma$). We also found one additional MW satellite candidate in the constellation of Horologium. This new object, Horologium\,II, was initially detected at the 7-8$\sigma$ significance levels in two separate, but overlapping DECam images. To fill the CCD chip gaps, we combined the photometric catalogs of the two frames and removed duplicates with $1\arcsec$ tolerance. Hor\,II was then recovered with a significance of $10\sigma$.

\begin{figure}
\begin{centering}
\includegraphics[scale=0.6]{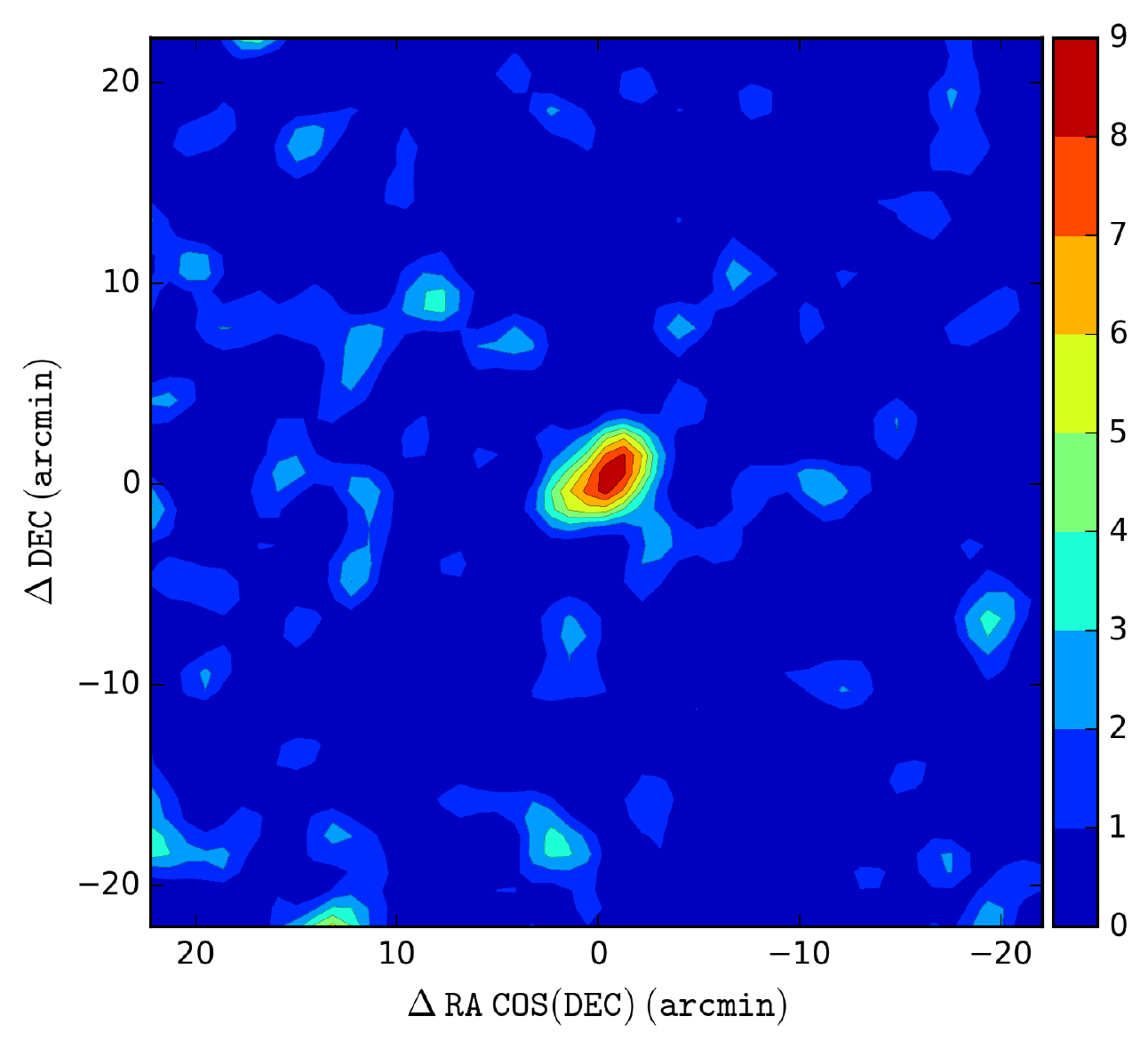}
\includegraphics[scale=0.6]{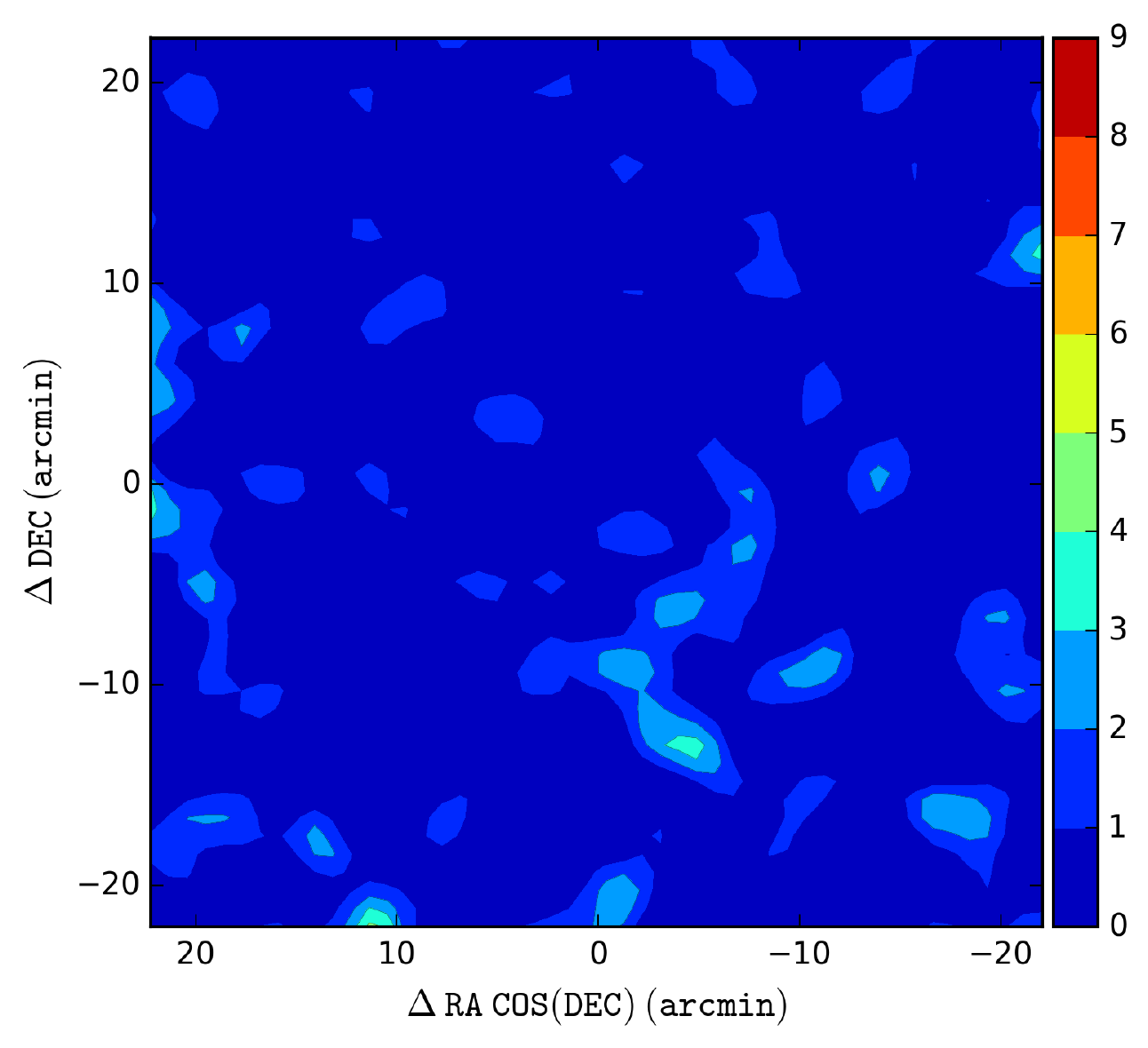}
\par\end{centering}
\caption{Upper panel: Smoothed density contour map of candidate stars, selected by the photometric filter with a PARSEC isochrone of 13.5 Gyr and [Fe/H]$=-2.1$~\citep{Parsec} shifted to the distance modulus $(m-M)=21.56$ magnitudes, centred on Hor\,II in the $42\times42$ square arcmin window. The contours represent the stellar density in units of the standard deviation above the background level. Lower panel: Same as the upper panel but for non-stellar objects, showing no overdensity consistent with Hor\,II in the upper panel. }
\label{fig:Contour}
\end{figure}

\begin{deluxetable}{lrl}
\tablewidth{0pt}
\tablecaption{Properties of Horologium \,II}
\tablehead{
\colhead{Parameter} & 
\colhead{Value} &
\colhead{Unit}}
\startdata
$\alpha_{J2000}$ & 3 16 32.1$\pm$5.0 & h m s \\
$\delta_{J2000}$ & $-$50 01 05$\pm$5 & $^\circ$ $\arcmin$ $\arcsec$ \\
$l$ & 262.472 & deg\\
$b$ & $-$54.137 & deg\\
$(m-M)$ & $19.46\pm0.20$ & mag \\
$d_\odot$ & $78\pm8$ & kpc \\
$r_{h}$ & $2.09^{+0.44}_{-0.41}$ & \arcmin \\
& $47\pm10$ \tablenotemark{a} & pc \\
$\epsilon$ & $0.52^{+0.13}_{-0.17}$ & \\
$\theta$ & $127\pm11$ & deg \\
$M_{V}$& $-2.6^{+0.2}_{-0.3} \CH{\tablenotemark{a}}$ & mag 
\enddata
\tablenotetext{a}{ Adopting a distance of 78\,kpc}
\label{tab:Parameters}
\end{deluxetable}

\begin{figure*}[t!]
\includegraphics[scale=0.73]{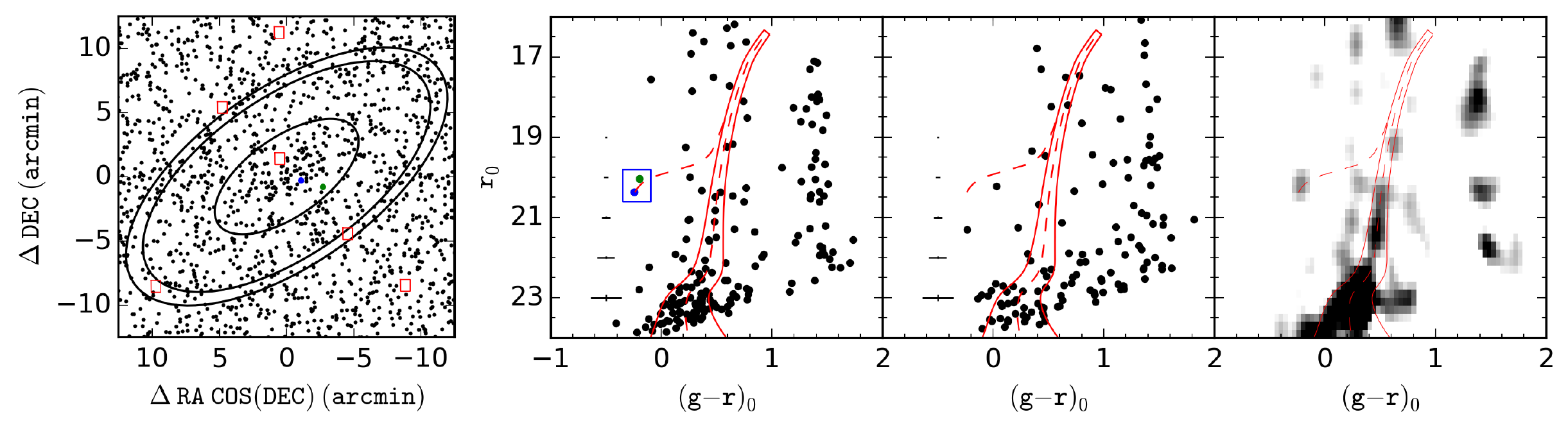}
\caption{DECam view of Hor\,II. Left panel: distribution of all stars in a $25' \times25'$ field centred on Hor\,II. The blue and green dots mark the two BHB candidates falling into the blue box in the next panel. The ellipses mark elliptical radii of $6\farcm3$ ($\sim 3r_{h}$), $12\farcm6$($\sim 6r_{h}$) and $14\farcm1$ respectively. The red rectangles mark the locations of residual CCD chip gap where none of DES Y1A1 images covers. Middle left panel: CMD of stars lying in the inner circle. Overplotted is the best-fitting PARSEC (red dashed lines) isochrone of age 13.5 Gyr and [Fe/H]=-2.1 shifted to the distance modulus of $(m-M)=19.46$. Within the isochrone mask (red solid lines) based on photometric uncertainties, there are candidate stars consistently aligned with the isochrone from the RGB to the main sequence turn-off. The two BHB candidates are colored and correspond to those in the left panel. Middle right panel: control CMD of stars falling in the annulus between the two outer circles but covering the same on-sky area as in the middle panel. Right panel: differential Hess diagram, the inner CMD minus the control CMD, which exhibits a clear excess of stars in the RGB and the main sequence turn-off regions. \label{fig:CMD}}
\end{figure*}

\begin{figure*}[t!]
\begin{centering}
\includegraphics[scale=0.83]{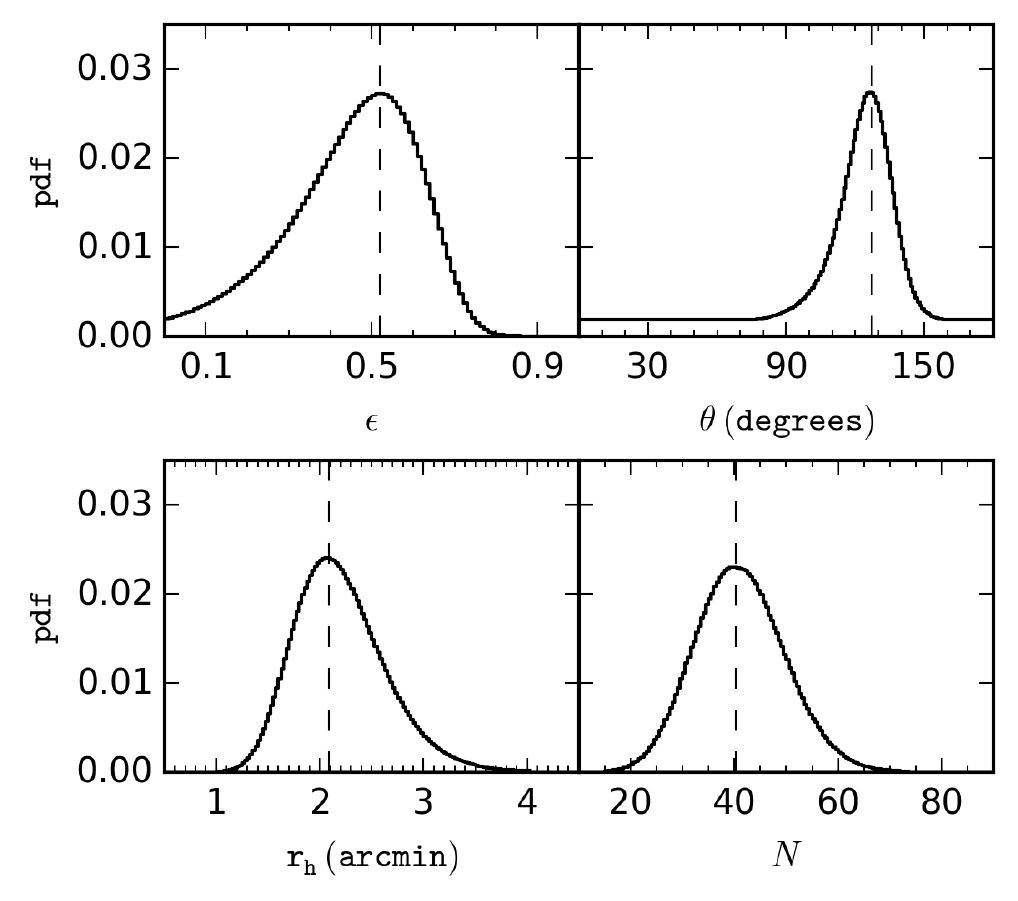}
\includegraphics[scale=0.83]{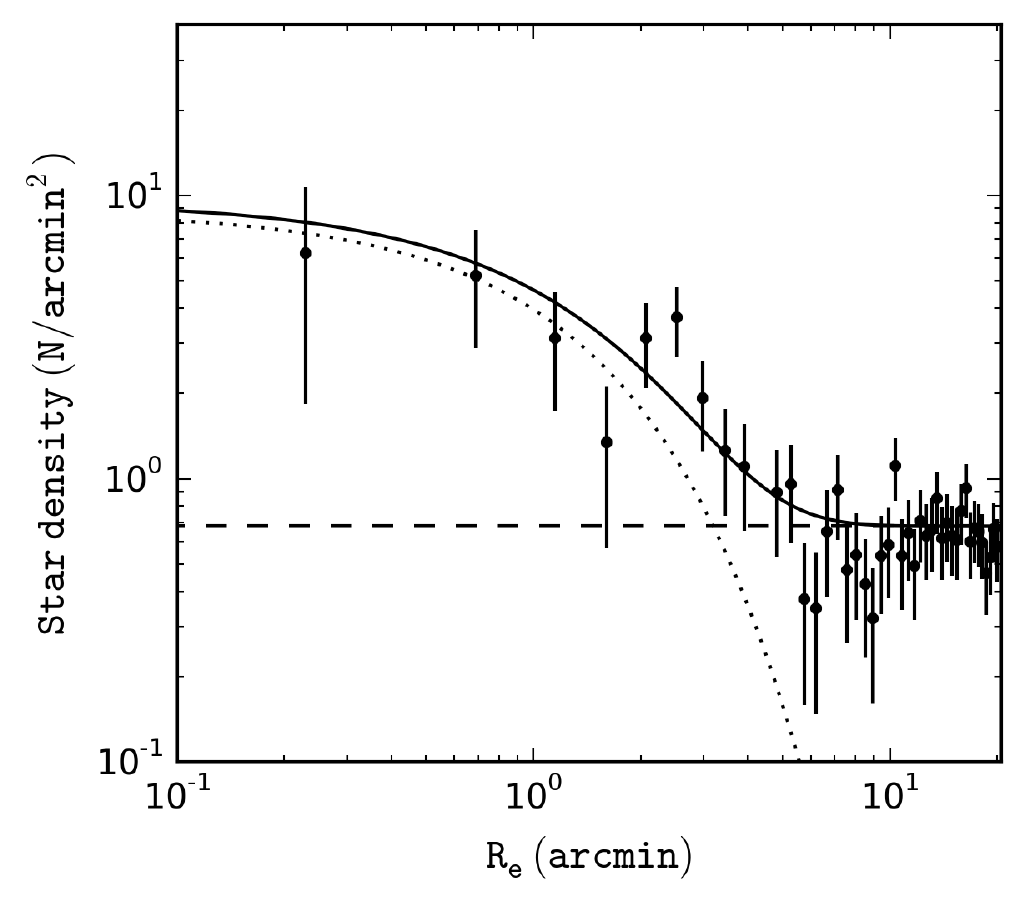}
\par\end{centering}
\protect\caption{Left panels: marginalized probability distribution functions for the the ellipticity  ($\epsilon$),  the position angle ($\theta$), they half-light radius ($r_{h}$) and the number of stars in the system for our photometric selection ($N$). The dashed lines mark the modes of the distributions. Right panel: radial density profile of Hor\,II, constructed on the mode values of each parameter as a function of elliptical radius $R_{e}$. The error of each data point is based on Poisson statistics. The dotted line marks the best-fit exponential model, the dashed line the foreground level and the solid line the combined fit. \label{fig:RadialProfile}}
\end{figure*}

The upper panel of Figure~\ref{fig:Contour} presents the density contour map of Hor\,II, made from stars passing the isochrone filter. Hor\,II appears elongated but well defined by high level density contours ($>3\sigma$). The lower panel shows the corresponding contour map of non-stellar objects in the same field of view.

\section{Candidate Properties}
The left panel of Figure~\ref{fig:CMD} shows the distribution of all stellar objects in our photometric catalogs within a $25\arcmin\times25\arcmin$ window centred on Hor\,II. The small red rectangles indicate the six locations of residual CCD chip gaps where the two DES images provided no data. In the middle panel, we present the color-magnitude diagram (CMD) for the stars in the inner circle shown in the left panel, equal to the dashed circle in the upper panel of Figure~\ref{fig:Contour}. Overplotted is the PARSEC isochrone~\citep{Parsec} of 13.5 Gyr and  [Fe/H]=-2.1 shifted to the distance modulus of $(m-M)=19.46\pm0.20$ or a heliocentric distance of $78\pm8$\,kpc. Compared to the control CMD established in the same manner as in~\cite{Belokurov2006} shown in the right panel, the stars in the vicinity of Hor\,II consistently trace the old and metal-poor stellar population from the red-giant branch (RGB) down to the main-sequence turn off. Hor\,II also hosts two potential blue horizontal branch (BHB) stars.

We derived the structural parameters of Hor\,II using a Maximum Likelihood (ML) algorithm similar to the one described in~\cite{Martin2008}. The resulting marginalized pdfs for the structural parameters are presented in the left panels of Figure~\ref{fig:RadialProfile}. The right panel shows the radial density profile with the best-fit exponential profile based on the best parameters; an ellipticity of $\epsilon=0.52$, a position angle of $\theta=127^{\circ}$ and a half-light radius of $r_{h}=2\farcm09$. Adopting the heliocentric distance of 78\,kpc, the ML-estimated physical size of Hor\,II is $r_{h}=47\pm10$\,pc.

The total luminosity of Hor\,II is estimated as follows. Briefly, we use the total number of member stars $N$ above the photometric threshold ($r_{0}\sim23.5$\,mag) and its associated uncertainty derived from the ML algorithm run. We then integrate a normalised theoretical luminosity function as a probability density function of magnitude, by the same magnitude limit and use the ratio of the number $N$ to the probability density to scale the luminosity function up to the observed level. Integrating the scaled luminosity function inclusive of missing flux below the threshold gives the absolute luminosity of Hor\,II. Using the PARSEC isochrone of  13.5 Gyr and  [Fe/H]=-2.1 based on the initial mass function by \cite{Kroupa2001}, we obtain $M_{r}\sim-2.74$ or $M_{V}\sim-2.57$ by the luminosity weighted mean color $V-r=0.17$ of the isochrone. We adopt a total luminosity of $M_{V}=-2.6^{+0.2}_{-0.3}$ as our final estimate where its uncertainty is derived from the star counts $N$. All the resulting parameters are summarised in Table~\ref{tab:Parameters}. We note that a heliocentric distance of 78\,kpc was adopted in the calculations of the physical size and total luminosity, to which the distance uncertainty was not propagated.

\section{Discussion and Conclusion}
We analysed the first instalment (Y1A1) of the Dark Energy Survey $gr$ imaging data to search for MW satellites, where we recovered all the previously reported systems and also found a new satellite candidate in the constellation of Horologium. The new MW satellite candidate Hor\,II appears faint ($M_{V}=-2.6^{+0.2}_{-0.3}$), elongated ($\epsilon=0.52^{+0.13}_{-0.17}$) and rather extended ($r_{h}=47\pm10$\,pc). On the size luminosity plane, Hor\,II is placed in the realm of UFDs close to Bo\"otes\,II. It also features a typically old ($\sim13.5$Gyr) and metal-poor ([Fe/H]$\sim-2.1$) stellar population. The best isochrone fit yields a heliocentric distance of $78\pm8$\,kpc, which is the same as that of a recently discovered neighbour in the DES Y1A1 coverage, Hor\,I. Compared to the new satellite candidate, Hor\,I is about twice as luminous ($M_{V}=-3.4\pm0.3$), smaller ($r_{h}=30^{+4.4}_{-3.3}$\,pc) and more circular ($\epsilon<0.28$), consequently being placed in the somewhat ambiguous region on the size-luminosity plane where UFDs and extended globular clusters overlap~\citep{Koposov2015a}. A recent spectroscopic study by \cite{Koposov2015b} has revealed that the dynamical mass-to-light ratio of Hor\,I reaches $M/L_V=570^{+1154}_{-112}$ and confirmed that that system is indeed an UFD, possibly (once) associated with the LMC. The pair of UFDs, Hor\,I and II, are only $\sim7$ degrees away from each other on the sky, have identical distances and are well aligned with the Vast Polar Structure~\citep{VPOS2015}. Such UFD pairs have been reported for quite some time e.g. Bo\"otes I - II~\citep{Walsh2007}, Leo IV - V~\citep{Belokurov2008} and Pisces II - Pegasus III~\citep{Kim2015b}. The tentative link between Hor\,I and II can be extended further to 
the Sculptor dwarf spheroidal~\citep[$d_\odot\sim82$\,kpc,][]{Weisz2014} and the UFD Eridanus\,III ($d_\odot=87\pm8$\,kpc). As Sculptor is moving on a retrograde orbit within the Vast Polar Structure when compared to the other classical MW satellite galaxies in the vicinity, including the Magellanic coulds, the hypothesis of a common origin can be tested once proper motion measurements become available.
 Nevertheless, this alignment is already suggestive of a ``layer" of outer halo UFDs parallel to the Magellanic Clouds, possibly associated to the most luminous Sculptor dwarf. Indeed, there is one more object nearby, namely Phe\,II, that also shares the same distance.

\acknowledgements{We thank the anonymous referee for the helpful comments and suggestions, which contributed to improving the quality of the publication. We acknowledge the support of the Australian Research Council through Discovery project DP150100862.

This paper makes use of data from the AAVSO Photometric All Sky Survey, whose funding has been provided by the Robert Martin Ayers Sciences Fund. This research made use of Astropy, a community-developed core Python package for Astronomy~\citep{astropy}, and Matplotlib library~\citep{matplotlib}.

This project used public archival data from the Dark Energy Survey (DES). Funding for the DES Projects has been provided by 
the U.S. Department of Energy, 
the U.S. National Science Foundation, 
the Ministry of Science and Education of Spain, 
the Science and Technology Facilities Council of the United Kingdom, 
the Higher Education Funding Council for England, 
the National Center for Supercomputing Applications at the University of Illinois at Urbana-Champaign, 
the Kavli Institute of Cosmological Physics at the University of Chicago, 
the Center for Cosmology and Astro-Particle Physics at the Ohio State University, 
the Mitchell Institute for Fundamental Physics and Astronomy at Texas A\&M University, 
Financiadora de Estudos e Projetos, Funda{\c c}{\~a}o Carlos Chagas Filho de Amparo {\`a} Pesquisa do Estado do Rio de Janeiro, 
Conselho Nacional de Desenvolvimento Cient{\'i}fico e Tecnol{\'o}gico and the Minist{\'e}rio da Ci{\^e}ncia, Tecnologia e Inovac{\~a}o, 
the Deutsche Forschungsgemeinschaft, 
and the Collaborating Institutions in the Dark Energy Survey. 
The Collaborating Institutions are 
Argonne National Laboratory, 
the University of California at Santa Cruz, 
the University of Cambridge, 
Centro de Investigaciones En{\'e}rgeticas, Medioambientales y Tecnol{\'o}gicas-Madrid, 
the University of Chicago, 
University College London, 
the DES-Brazil Consortium, 
the University of Edinburgh, 
the Eidgen{\"o}ssische Technische Hoch\-schule (ETH) Z{\"u}rich, 
Fermi National Accelerator Laboratory, 
the University of Illinois at Urbana-Champaign, 
the Institut de Ci{\`e}ncies de l'Espai (IEEC/CSIC), 
the Institut de F{\'i}sica d'Altes Energies, 
Lawrence Berkeley National Laboratory, 
the Ludwig-Maximilians Universit{\"a}t M{\"u}nchen and the associated Excellence Cluster Universe, 
the University of Michigan, 
{the} National Optical Astronomy Observatory, 
the University of Nottingham, 
the Ohio State University, 
the University of Pennsylvania, 
the University of Portsmouth, 
SLAC National Accelerator Laboratory, 
Stanford University, 
the University of Sussex, 
and Texas A\&M University.

}

%\bibliographystyle{apj}
%\bibliography{paper}

\begin{thebibliography}{}

\bibitem[Astropy Collaboration et 
al.(2013)]{astropy} Astropy Collaboration, Robitaille, T.~P., Tollerud, E.~J., et al.\ 2013, \aap, 558, A33 

\bibitem[Bechtol et al.(2015)]{Bechtol2015} Bechtol, K., Drlica-Wagner, A., et al.\ 2015, 
arXiv:1503.02584 

\bibitem[Belokurov et al.(2006)]{Belokurov2006} Belokurov, V., 
Zucker, D.~B., Evans, N.~W., et al.\ 2006, \apjl, 647, L111

\bibitem[Belokurov et al.(2008)]{Belokurov2008} Belokurov, V., 
Walker, M.~G., Evans, N.~W., et al.\ 2008, \apjl, 686, L83 

\bibitem[Belokurov et al.(2014)]{Belokurov2014} Belokurov, V., Irwin, 
M.~J., Koposov, S.~E., et al.\ 2014, \mnras, 441, 2124 

\bibitem[Bertin 
\& Arnouts(1996)]{SExtractor} Bertin, E., \& Arnouts, S.\ 1996, \aaps, 117, 393 

\bibitem[Bertin(2011)]{PSFEx} Bertin, E.\ 2011, Astronomical 
Data Analysis Software and Systems XX, 442, 435 

\bibitem[Bressan et al.(2012)]{Parsec} Bressan, A., Marigo, 
P., Girardi, L., et al.\ 2012, \mnras, 427, 127 

\bibitem[Brown et al.(2014)]{Brown2014} Brown, T.~M., Tumlinson, 
J., Geha, M., et al.\ 2014, \apj, 796, 91 

\bibitem[Desai et al.(2012)]{Desai2012} Desai, S., Armstrong, R., 
Mohr, J.~J., et al.\ 2012, \apj, 757, 83 

\bibitem[Frebel et al.(2010)]{Frebel2010} Frebel, A., Simon, 
J.~D., Geha, M., \& Willman, B.\ 2010, \apj, 708, 560 

\bibitem[Grillmair(2009)]{Grillmair2009} Grillmair, C.~J.\ 2009, 
\apj, 693, 1118 

\bibitem[Hunter(2007)]{matplotlib} Hunter, J.~D.\ 2007, Computing 
in Science and Engineering, 9, 90 

\bibitem[Irwin et al.(2007)]{Irwin2007} Irwin, M.~J., Belokurov, 
V., Evans, N.~W., et al.\ 2007, \apjl, 656, L13 

\bibitem[Jerjen(2010)]{SMS} Jerjen, H.\ 2010, Advances in 
Astronomy, 2010, 434390

\bibitem[Kim 
\& Jerjen(2015)]{Kim1} Kim, D., \& Jerjen, H.\ 2015, \apj, 799, 73 

\bibitem[Kim et al.(2015a)]{Kim2015a} Kim, D., Jerjen, H., 
Milone, A.~P., Mackey, D., \& Da Costa, G.~S.\ 2015, \apj, 803, 63

\bibitem[Kim et al.(2015b)]{Kim2015b} Kim, D., Jerjen, H., 
Mackey, D., Da Costa, G.~S., \& Milone, A.~P.\ 2015, arXiv:1503.08268  

\bibitem[Kirby et al.(2008)]{Kirby2008} Kirby, E.~N., Simon, 
J.~D., Geha, M., Guhathakurta, P., \& Frebel, A.\ 2008, \apjl, 685, L43 

\bibitem[Kroupa(2001)]{Kroupa2001} Kroupa, P.\ 2001, \mnras, 322, 
231 

\bibitem[Koposov et al.(2008)]{Koposov2008} Koposov, S., Belokurov, 
V., Evans, N.~W., et al.\ 2008, \apj, 686, 279 

\bibitem[Koposov et al.(2015a)]{Koposov2015a} Koposov, S.~E., 
Belokurov, V., Torrealba, G., \& Wyn Evans, N.\ 2015, arXiv:1503.02079 

\bibitem[Koposov et al.(2015b)]{Koposov2015b} Koposov, S.~E., Casey, 
A.~R., Belokurov, V., et al.\ 2015, arXiv:1504.07916 

\bibitem[Laevens et al.(2014)]{Laevens2014} Laevens, B.~P.~M., 
Martin, N.~F., Sesar, B., et al.\ 2014, \apjl, 786, L3 

\bibitem[Laevens et al.(2015)]{Laevens2015} Laevens, B.~P.~M., 
Martin, N.~F., Ibata, R.~A., et al.\ 2015, arXiv:1503.05554

\bibitem[Martin et al.(2007)]{Martin2007} Martin, N.~F., Ibata, 
R.~A., Chapman, S.~C., Irwin, M., \& Lewis, G.~F.\ 2007, \mnras, 380, 281 

\bibitem[Martin et al.(2008)]{Martin2008} Martin, N.~F., de Jong, 
J.~T.~A., \& Rix, H.-W.\ 2008, \apj, 684, 1075

\bibitem[Martin et al.(2015)]{Martin2015} Martin, N.~F., Nidever, 
D.~L., Besla, G., et al.\ 2015, \apjl, 804, L5 

\bibitem[Mohr et al.(2012)]{DESDM} Mohr, J.~J., Armstrong, 
R., Bertin, E., et al.\ 2012, \procspie, 8451, 84510D 

\bibitem[Mu{\~n}oz et al.(2010)]{Munoz2010} Mu{\~n}oz, R.~R., 
Geha, M., \& Willman, B.\ 2010, \aj, 140, 138 

\bibitem[Norris et al.(2010)]{Norris2010} Norris, J.~E., Yong, D., 
Gilmore, G., \& Wyse, R.~F.~G.\ 2010, \apj, 711, 350

\bibitem[Pawlowski et al.(2015)]{VPOS2015} Pawlowski, M.~S., 
McGaugh, S.~S., \& Jerjen, H.\ 2015, arXiv:1505.07465 

\bibitem[Sand et al.(2012)]{Sand2012} Sand, D.~J., Strader, J., 
Willman, B., et al.\ 2012, \apj, 756, 79 

\bibitem[Shanks et al.(2015)]{Shanks2015} Shanks, T., Metcalfe, 
N., Chehade, B., et al.\ 2015, arXiv:1502.05432 

\bibitem[Schlafly 
\& Finkbeiner(2011)]{Schlafly2011} Schlafly, E.~F., \& Finkbeiner, D.~P.\ 2011, \apj, 737, 103 

\bibitem[Schlegel et al.(1998)]{Schlegel1998} Schlegel, D.~J., 
Finkbeiner, D.~P., \& Davis, M.\ 1998, \apj, 500, 525 

\bibitem[Simon \& Geha(2007)]{Simon2007} Simon, J.~D., \& Geha, M.\ 2007, \apj, 670, 313

\bibitem[Simon et al.(2011)]{Simon2011} Simon, J.~D., Geha, M., 
Minor, Q.~E., et al.\ 2011, \apj, 733, 46 

\bibitem[Simon et al.(2015)]{Simon2015} Simon, J.~D., 
Drlica-Wagner, A., Li, T.~S., et al.\ 2015, arXiv:1504.02889

\bibitem[Taylor(2005)]{STILTS} Taylor, M.~B.\ 2005, 
Astronomical Data Analysis Software and Systems XIV, 347, 29 

\bibitem[The Dark Energy Survey Collaboration(2005)]{DES} 
The Dark Energy Survey Collaboration 2005, arXiv:astro-ph/0510346 

\bibitem[Walker et al.(2015)]{Walker2015} Walker, M.~G., Mateo, 
M., Olszewski, E.~W., et al.\ 2015, arXiv:1504.03060 

\bibitem[Walsh et al.(2007)]{Walsh2007} Walsh, S.~M., Jerjen, H., 
\& Willman, B.\ 2007, \apjl, 662, L83 

\bibitem[Walsh et al.(2009)]{Invisibles} Walsh, S.~M., Willman, 
B., \& Jerjen, H.\ 2009, \aj, 137, 450 

\bibitem[Weisz et al.(2014)]{Weisz2014} Weisz, D.~R., Dolphin, 
A.~E., Skillman, E.~D., et al.\ 2014, \apj, 789, 147

\bibitem[Willman et al.(2005)]{Willman2005} Willman, B., Blanton, 
M.~R., West, A.~A., et al.\ 2005, \aj, 129, 2692 

\bibitem[Willman 
\& Strader(2012)]{Willman2012} Willman, B., \& Strader, J.\ 2012, \aj, 144, 76 

\bibitem[York et al.(2000)]{York2000} York, D.~G., Adelman, J., 
Anderson, J.~E., Jr., et al.\ 2000, \aj, 120, 1579 

\bibitem[Zucker et al.(2006)]{Zucker2006} Zucker, D.~B., 
Belokurov, V., Evans, N.~W., et al.\ 2006, \apjl, 643, L103 

\end{thebibliography}

\end{document}